\documentclass[reqno]{amsart}
\usepackage{amsbsy,eucal,style}

\theoremstyle{plain}
\newtheorem{theorem}{Theorem}[section]
\newtheorem{proposition}[theorem]{Proposition}
\newtheorem{corollary}[theorem]{Corollary}
\newtheorem{lemma}[theorem]{Lemma}
\newenvironment{odstavec}{\refstepcounter{theorem}
                          \vspace{2ex}\noindent
                          {\bf\thetheorem.}\rm }
                          {\mbox{}\hfill\vspace{2ex}}

\theoremstyle{definition}
\newtheorem{definition}[theorem]{Definition}
\newtheorem{example}[theorem]{Example}
\newtheorem{examples}[theorem]{Examples}

\newtheorem{remark}[theorem]{Remark}
\newtheorem{notation}[theorem]{Notation}
\newtheorem{observation}[theorem]{Observation}
\newenvironment{remarkNoNumber}{\vspace{2ex}\noindent
                          {\bf{Remark.}}\ \rm }
                          {\mbox{}\hfill\vspace{2ex}}

\input xy.tex
\xyoption{matrix} 
\xyoption{arrow}
\xyoption{curve}

\numberwithin{equation}{section}

\begin{document}
\title[On Coalgebra Based on Classes]
      {On Coalgebra Based on Classes}
\author[Ji\v{r}\'{\i} Ad\'{a}mek]{Ji\v{r}\'{\i} Ad\'{a}mek${}^{*)}$}
\author{Stefan Milius}
\author[Ji\v{r}\'{\i} Velebil]{Ji\v{r}\'{\i} Velebil${}^{*)}$}
\address{Institute of Theoretical Computer Science, Technical University,
         Braunschweig, Germany}
\email{$\{$adamek,milius,velebil$\}$@iti.cs.tu-bs.de}
\thanks{${}^{*)}$The first and the third author acknowledge the support
        of the Grant Agency of the Czech
        Republic under the Grant No.~201/02/0148.}
\date{January 22, 2003}

\begin{abstract}
The category $\Class$ of classes and functions is proved to have
a number of properties suitable for algebra and coalgebra: every 
endofunctor has an initial algebra and a terminal coalgebra, the 
categories of algebras and coalgebras are complete and cocomplete,
and every endofunctor generates a free completely iterative monad.
A description of a terminal coalgebra for the power-set functor
is provided.
\end{abstract}

\maketitle

\section{Introduction: Don't Be Afraid of Classes}
This paper does not, despite its title, concern the foundations.
It concerns coalgebra in a surprisingly coalgebra-friendly category
\begin{displaymath}
\Class
\end{displaymath}
of classes and functions --- and the main message is that one almost
does not need foundations for that, or just the reasonable minimum
of foundations. What is the definition of reasonable minimum? In 
category theory one always works with ``large'' and ``small'' --- and
this is all one needs. Thus, ``large'' refers to, say, set theory
which is a model of ${\sf{ZFC}}$ (Zermelo-Fraenkel axioms including
the Axiom of Choice). And ``small'' means that a universe (of small
sets) is once for all chosen within the given universe (of all sets).
This is all one needs: the chosen universe of all small sets is itself
a large set, and we denote by
\begin{displaymath}
\Aleph
\end{displaymath}
its cardinality. This means that the category $\Set$ of all small
sets is, obviously, equivalent to the category $\SET_{<\Aleph}$
of all large sets of cardinality smaller than $\Aleph$. And if one 
forms, analogously, the category $\SET_{\leq\Aleph}$ of all large
sets of cardinality less or equal to $\Aleph$, then $\Class$ is
equivalent to it. Thus, one can think of the difference between
$\Set$ and $\Class$ as of the difference between being smaller than,
or smaller or equal to, $\Aleph$. The cardinal $\Aleph$ is strongly
inaccessible (i.e., for every cardinal $\alpha < \Aleph$ we have
$2^\alpha < \Aleph$), and conversely, for every choice of a strongly
inaccessible uncountable cardinal $\Aleph$ there is a universe of 
small sets with $\Set\simeq\SET_{<\Aleph}$.

In what follows we work with the category of all sets of cardinality
less than $\Aleph$ as $\Set$ and with the category 
of all sets of cardinality at most $\Aleph$ as the category $\Class$.

\section{All Endofunctors are Set-Based}
\label{sec:2}

The concept of a set-based endofunctor of $\Class$ has been introduced
by Peter Aczel and Nax Mendler~\cite{am} in order to prove their
``Final Coalgebra Theorem'', see the next section. An endofunctor
$F:\Class\to\Class$ is called {\em set-based} provided that for every class
$X$ and every element of $FX$ there exists a small subset $m:M\subto X$
such that that element lies in the image of $Fm$. It turns out that 
every endofunctor has this property. The following proof, based on ideas
of V\'{a}clav Koubek~\cite{k}, uses a classical set-theoretical result
of Alfred Tarski; see~\cite{t}:

\begin{theorem}
\label{th:2.2}
For every infinite cardinal $\lambda$ there exists, on a set $X$
of cardinality $\lambda$, an almost disjoint collection of subsets
$X_i\subseteq X$, $i\in I$, i.e., a collection satisfying
${\mathrm{card}}\; (X_i\cap X_j)<\lambda$ for all $i\not= j$ in $I$,
such that $I$ has more than $\lambda$ elements.
\end{theorem}

\begin{theorem}
\label{th:2.3}
Every endofunctor of $\Class$ is set-based.
\end{theorem}

\begin{remarkNoNumber}
We prove a more general statement: given an infinite
regular
\footnote{Recall that $\lambda$ is regular if it is not a sum of less
than $\lambda$ smaller cardinals.}
cardinal $\lambda$, then every endofunctor of $\SET_{\leq\lambda}$
(the category of all sets of cardinality at most $\lambda$) is 
$\lambda$-accessible. Recall from~\cite{ap1} that for an endofunctor
$F$ of $\SET_{\leq\lambda}$ the following conditions are equivalent:
\begin{enumerate}
\item $F$ is $\lambda$-accessible, i.e., preserves $\lambda$-filtered
      colimits.
\item For every set $X$ in $\SET_{\leq\lambda}$ and every element
      $x$ of $FX$ there exists a subset $m:M\subto X$ of cardinality
      less than $\lambda$ such that $x$ lies in the image
      of $Fm$.
\item $F$ is a quotient of a $\lambda$-ary polynomial functor.
\end{enumerate}
(Given a signature $\Sigma$, i.e., a set of operation symbols
$\sigma$ with prescribed arities ${\mathrm{ar}}(\sigma)$, finite
or infinite, then the corresponding functor
\begin{displaymath}
H_\Sigma:X\mapsto\coprod_{\sigma\in\Sigma} X^{{\mathrm{ar}}(\sigma)}
\end{displaymath}
is called {\em polynomial}. It is {\em $\lambda$-ary} if 
${\mathrm{ar}}(\sigma)<\lambda$ for all $\sigma\in\Sigma$.)

In particular, set-based and $\Aleph$-accessible are equivalent.
\end{remarkNoNumber}

\begin{proof}[Proof of Theorem and Remark]
Let $\lambda$ be an infinite cardinal. Given 
$F:\SET_{\leq\lambda}\to\SET_{\leq\lambda}$ and a set
$X$ in $\SET_{\leq\lambda}$, then for every element $x\in FX$ 
we are to find a subset $m:M\subto X$ with ${\mathrm{card}}\; M<\lambda$
and $x\in Fm[FM]$. If ${\mathrm{card}}\; X<\lambda$ there is nothing
to prove, assume ${\mathrm{card}}\; X=\lambda$. We can further assume, 
without loss of generality, that $F$ preserves finite intersections.
In fact, by a theorem of V\v{e}ra Trnkov\'{a}, see, e.g., Theorem~III.4.5 
of~\cite{at}, there exists a functor
$F'$ preserving finite intersections and such that the restrictions
of $F$ and $F'$ to the full subcategory $\Set_{\leq\lambda}$
of all nonempty sets are naturally isomorphic. Since $F$ is 
$\lambda$-accessible iff $F'$ is, we can assume $F'=F$.
By Theorem~\ref{th:2.2} there exists an almost disjoint collection
of subsets $v_i:X_i\subto X$, $i\in I$, with ${\mathrm{card}}\; I>\lambda$.
Since the collection of all subsets of $X$ of cardinality less than
$\lambda$ has cardinality $\lambda$ (due to the regularity of $\lambda$),
we can suppose without loss of generality that each $X_i$ has cardinality
$\lambda$ --- in fact, by discarding all $X_i$ of cardinalities less
than $\lambda$ we still obtain an almost disjoint collection of more
than $\lambda$ members. For each $i\in I$ thus there exists an 
isomorphism
\begin{displaymath}
w_i:X\to X_i
\end{displaymath}
and we put
\begin{displaymath}
y_i=F(v_i w_i)(x)\in FX.
\end{displaymath}
Since $F$ is an endofunctor of $\SET_{\leq\lambda}$, the set
$FX$ has cardinality smaller than that of $I$, consequently, 
the elements $y_i$ are not pairwise disjoint. Choose $i\not= j$
in $I$ with 
\begin{displaymath}
y_i = y_j
\end{displaymath}
and form a pullback (an intersection of $v_i$ and $v_j$):
\begin{displaymath}
\xy
\xymatrix{
&
M
\ar[1,-1]_{u_i}
\ar[1,1]^{u_j}
&
\\
X_i
\ar[1,1]_{v_i}
&
&
X_j
\ar[1,-1]^{v_j}
\\
&
X
&
}
\endxy
\end{displaymath}
Since $F$ preserves this pullback and $Fv_i(Fw_i(x))=Fv_j(Fw_j(x))$,
there exists
\begin{displaymath}
y\in FY
\quad
\mbox{with}
\quad
Fu_i(y)=Fw_i(x).
\end{displaymath}
For the subobject $m=w_i^{-1}u_i:M\subto X$ this implies 
\begin{displaymath}
Fm(y)=x
\end{displaymath}
and this concludes the proof.  
\end{proof}

\begin{remark}
\label{rem:extension}
Denote by $J:\Set\to\Class$ the inclusion functor. That
a functor $F:\Class\to\Class$ is set-based can be equivalently
restated as being naturally isomorphic to a left Kan extension
$\Lan{J}{K}$ for some functor $K:\Set\to\Class$.

Thus, Theorem~\ref{th:2.3} says that restriction 
along $J$, i.e., the functor
\begin{displaymath}
\,\_\,\o J : [\Class,\Class]\to [\Set,\Class]
\end{displaymath} 
is an equivalence of categories. We denote the pseudoinverse
of this restriction by $(\,\_\,)^\sharp$.

In fact, $J:\Set\to\Class$ is a free cocompletion of
$\Set$ under colimits of transfinite chains, i.e., colimits of chains 
indexed by the set of all small ordinals (see~\cite{amv}).
Thus, every functor of the form $F^\sharp$ (i.e., {\em every}
endofunctor of $\Class$) preserves colimits of transfinite chains.
\end{remark}

\begin{notation}
For an endofunctor $F:\Set\to\Set$ we denote by
\begin{displaymath}
F^\infty:\Class\to\Class
\end{displaymath}
the extension $(J\o F)^\sharp$ of the composite $J\o F:\Set\to\Class$.
\end{notation}

\section{All Functors are Varietors and Covarietors}

In the present section we show that endofunctors $H$ of $\Class$
have a surprisingly simple structure, and they admit free $H$-algebras
(i.e., are varietors) and cofree $H$-coalgebras 
(i.e., are covarietors) --- moreover, these algebras and coalgebras
can be explicitly described.

\begin{odstavec}
{\bf Polynomial Endofunctors.}
Classical Universal Algebra deals with $\Sigma$-algebras in the
category $\Set$, where $\Sigma$ is a (small) signature, i.e., 
a small set of operation symbols $\sigma$ with prescribed arities
${\mathrm{ar}}(\sigma)$ which are (in general, infinite) small 
cardinal numbers. Thus, if
\begin{displaymath}
\Card
\end{displaymath}
denotes the class of all small cardinal numbers, then a small
signature is a small set $\Sigma$ equipped with a function
${\mathrm{ar}}:\Sigma\to\Card$. And $\Sigma$-algebras are just
algebras over the polynomial endofunctor $H_\Sigma$ of $\Set$
given on objects, $X$, by
\begin{displaymath}
H_\Sigma X = \coprod_{\sigma\in\Sigma} X^{{\mathrm{ar}}(\sigma)}
\end{displaymath}
Quite analogously, in $\Class$ we work with (large) {\em signatures}
as classes $\Sigma$ equipped with a function
${\mathrm{ar}}:\Sigma\to\Card$ (thus, largeness refers to the 
possibility of having a proper class of operations, arities are
small). Here, again, we obtain a {\em polynomial endofunctor}
$H_\Sigma$ defined on classes $X$ by 
\begin{displaymath}
H_\Sigma X = \coprod_{\sigma\in\Sigma} X^{{\mathrm{ar}}(\sigma)}
\end{displaymath}
and analogously on morphisms.
\end{odstavec}

\begin{proposition}
\label{prop:3.3}
Every endofunctor $H$ of $\Class$ is a quotient of a polynomial
functor. That is, there exists a natural epitransformation
$\eps:H_\Sigma\to H$ for some signature $\Sigma$.
\end{proposition}
\begin{proof}
Let $\Sigma$ be the signature which, for every small cardinal $n$
has as $n$-ary symbols precisely the elements of $Hn$. Then the 
function
\begin{displaymath}
\eps_X:\coprod_{n\in\Card}\coprod_{\sigma\in Hn} X^n\to HX
\end{displaymath}
which to every $f:n\to X$ in the $\sigma$-th summand $X^n$ assigns
$Hf(\sigma)$ in $HX$ is a component of a natural transformation
(due to Yoneda lemma). And $\eps$ is pointwise surjective: 
for a small set $M$ put $n={\mathrm{card}}\; M$ and choose an isomorphism 
$f:n\to M$. Then every element of $HM$ has the form $Hf(\sigma)$ 
for a unique $\sigma\in\Sigma$. Thus, $\eps_M$ is surjective.
For a general $X$ use the fact that
$H$ is set-based (Theorem~\ref{th:2.3}), thus for an element
$x\in HX$ there exists a small subset $m:M\subto X$ and $y\in HM$
such that $x=Hm(y)$. Since $\eps_M$ is surjective, there exists
$z\in H_\Sigma M$ such that $\eps_M(z)=y$. Define 
$t=H_\Sigma m(z)\in H_\Sigma X$. Due to naturality of $\eps$
it follows that $\eps_X(t)=x$.
\end{proof}

\begin{example}
The power-set functor $\P:\Set\to\Set$ extends uniquely to
$\P^\infty:\Class\to\Class$, see Remark~\ref{rem:extension}.
The functor $\P^\infty$ assigns to every class $X$ the class
of all small subsets of $X$.
We can represent $\P^\infty$ as a quotient of $H_{\Sigma^0}$ where
$\Sigma^0$ is the signature which possesses, for every cardinal
$n\in\Card$, a unique operation $\sigma_n$: here
\begin{displaymath}
\eps_X : H_{\Sigma^0} = \coprod_{n\in\Card} X^n \to \P^\infty X
\end{displaymath}
assigns to every $f:n\to X$ the image $f[n]\subseteq X$.
\end{example}

\begin{odstavec}
{\bf Algebras.}
Recall that for an endofunctor $H$ of $\Class$ an {\em $H$-algebra}
is a class $A$ together with a function $\alpha:HA\to A$.
Given another algebra $\beta:HB\to B$, a {\em homomorphism} 
from $A$ to $B$ is a function $f:A\to B$ such that the following 
square
\begin{displaymath}
\miniSQUARE{HA}
           {\alpha}
           {A}
           {Hf}
           {f}
           {HB}
           {\beta}
           {B}
\end{displaymath}
commutes. The category of all $H$-algebras and homomorphisms 
is denoted by
\begin{displaymath}
\Alg{H}
\end{displaymath}
\end{odstavec}

\begin{examples}
\mbox\hfill
\begin{enumerate}
\item $\Alg{H_\Sigma}$ is the category of $\Sigma$-algebras
      (i.e., classes $A$ endowed, for every $n$-ary symbol $\sigma$,
      with an $n$-ary operation on $A$) which, except for the
      ``size'' of underlying sets, is just the classical category
      of Universal Algebra. 
\item $\Alg{\P^\infty}$ has as objects classes $A$ together with
      a function $\alpha:\P^\infty A\to A$. This can be equivalently
      considered as a variety of $\Sigma^0$-algebras as follows:
      let $E$ be the class of all equations
      \begin{displaymath}
      \sigma_n(x_i)_{i<n}\approx \sigma_m(y_j)_{j<m}
      \end{displaymath}
      where $n$ and $m$ are small cardinals and the variables
      $x_i$ and $y_j$ are such that the sets
      $\{ x_i\mid i<n \}$ and $\{ y_j\mid j<m \}$
      are equal. Then $\Alg{\P^\infty}$ is isomorphic to the
      variety of all $\Sigma^0$-algebras satisfying the above 
      equations.
\end{enumerate}
\end{examples}

\begin{remark}
Recall from~\cite{at} that a {\em basic equation} is an equation
between two flat terms, i.e., terms of the form $\sigma(x_i)_{i<n}$
where $\sigma$ is an $n$-ary operation symbol and $x_i$ are 
(not necessarily distinct) variables. The example $\Alg{\P^\infty}$
above is quite typical: every category $\Alg{H}$ is a variety
presented by basic equations (and vice versa) --- this has been 
shown for $\Set$ in~\cite{at}, let us recall it and extend to
the present ambient:

Given a functor $H$ represented as in Proposition~\ref{prop:3.3},
consider all the basic equations
\begin{displaymath}
\sigma(x_i)_{i<n}\approx \rho(y_j)_{j<m}
\end{displaymath}
where $\sigma,\rho\in\Sigma$ and for the set 
$V=\{ x_i\mid i<n\}\cup\{ y_j\mid j<m\}$ of variables we have:
\begin{itemize}
\item[] $\eps_V$ merges the $n$-tuple $(x_i)_{i<n}$ in the 
        $\sigma$-summand of $H_\Sigma V$ with the $m$-tuple
        $(y_j)_{j<m}$ in the $\rho$-summand of $H_\Sigma V$.
\end{itemize}
Then $\Alg{H}$ is equivalent to the variety of all $\Sigma$-algebras
presented by the above equations.

Conversely, given a class $E$ of basic equations in signature
$\Sigma$, there is a quotient $H$ of $H_\Sigma$ such that the
variety of $\Sigma$-algebras presented by $E$ is isomorphic to
$\Alg{H}$.
\end{remark}

\begin{corollary}
\label{cor:3.7}
For every endofunctor $H$ of $\Class$ an initial $H$-algebra
exists.
\end{corollary}

In fact, we can describe an initial $H$-algebra, $I$, in two
substantially different ways:
\begin{enumerate}
\item $I$ is a quotient of the initial $\Sigma$-algebra 
      modulo the congruence generated by the given basic
      equations.

      \medskip\noindent
      Recall here the description of initial $\Sigma$-algebras
      well-known in Universal Algebra: it is the algebra of all
      well-founded $\Sigma$-trees. This remains unchanged in case
      of large signatures, the only difference is that all $\Sigma$-trees
      do not form a small set (but each $\Sigma$-tree is small, 
      by definition). That is, by a {\em $\Sigma$-tree} we mean an ordered, 
      labelled tree on a small set of nodes, where labels are operation 
      symbols, and every node labelled by an $n$-ary symbol has precisely 
      $n$ children. The algebra
      \begin{displaymath}
      I_\Sigma
      \end{displaymath} 
      of all {\em well-founded} $\Sigma$-trees, i.e., $\Sigma$-trees
      in which every branch is finite, has operations given 
      by tree-tupling. This is an initial algebra in $\Alg{H_\Sigma}$.

      Given a quotient $\eps:H_\Sigma\to H$, form the smallest congruence
      $\sim$ on $I_\Sigma$ which is generated by all the basic equations 
      corresponding to $\eps$. Then $I_\Sigma/\sim$ is an initial
      algebra of $\Alg{H}$.
\item $I$ is a colimit of the transfinite chain $W:\Ord\to\Class$
      (where $\Ord$ is the chain of all small ordinals) given by
      iterating $H$ on the initial object $\emptyset$ of $\Class$:
      \begin{itemize}
      \item[] $W_i=H^i(\emptyset)$
      \end{itemize}
      and
      \begin{itemize}
      \item[] $I=\colim_{i\in\Ord} W_i$.
      \end{itemize}

      \medskip\noindent
      More precisely, there is a unique chain $W$ for which we have
      \begin{description}
      \item[First step] $W_0=\emptyset$, $W_1=H(\emptyset)$ and
           $W_{0,1}:\emptyset\to H(\emptyset)$ unique.
      \item[Isolated step] $W_{i+1}=H(W_i)$ and 
           $W_{i+1,j+1}=H(W_{i,j})$.
      \item[Limit step] $W_j=\colim_{i<j} W_i$ with colimit 
           cocone $(W_{i,j})_{i<j}$.
      \end{description}
      A colimit of this chain exists (see Observation~\ref{obs:3.9}
      below) and is preserved by $H$, see~\ref{rem:extension}, therefore, if 
      $I=\colim W_i$ then
      \begin{displaymath}
      HI
      \cong
      \colim_{i\in\Ord} H(W_i)
      =
      \colim_{i\in\Ord} W_{i+1}
      \cong
      I 
      \end{displaymath}
      and the canonical isomorphism $HI\to I$ defines an initial
      $H$-algebra, see~\cite{a1}.
\end{enumerate}

\begin{example}
\label{ex:3.8}
An initial $\P^\infty$-algebra. Whereas the power-set functor $\P$
has no initial algebra, $\P^\infty$ does (since every endofunctor
of $\Class$ does). The above chain $W_i$ coincides with the chain 
of sets defined by the cumulative hierarchy:
\begin{itemize}
\item[] $W_0=\emptyset$
\item[] $W_{i+1}={\mathrm{exp}}\; W_i$
\end{itemize}
and
\begin{itemize}
\item[] $W_j=\displaystyle{\bigcup_{i<j}} W_i$ for limit ordinals $j$. 
\end{itemize}
Consequently, we can describe an initial $\P^\infty$-algebra
as
\begin{displaymath}
I
=
\Set
=
\mbox{the class of all small sets}
\end{displaymath}
with the structure map $\P^\infty I\to I$ given by the union.
This has been first observed by Jan Rutten and Danielle Turi~\cite{rt}.

The other option of describing $I$ is also interesting: let us first
form the initial $\Sigma^0$-algebra. Since operations of any arity
are unique, we can first forget the labelling, thus
\begin{displaymath}
I_{\Sigma^0}
=
\mbox{the algebra of all well-founded trees}
\end{displaymath}
To every well-founded tree $t$ let us assign the corresponding non-ordered 
tree (obtained by forgetting the linear ordering of children of any
node) and recall that a non-ordered tree is called {\em extensional}
provided that every pair of distinct siblings defines a pair of
non-isomorphic subtrees. For every $t\in I_{\Sigma^0}$ denote by
$[t]$ the {\em extensional quotient} of the (non-ordered version of)
$t$; that is, the extensional tree obtained from $t$ by iteratively
merging any pair of siblings defining isomorphic subtrees. Then an
initial $\P^\infty$-algebra can be described as
\begin{displaymath}
I_{\Sigma^0}/\sim
\quad
\mbox{where $t\sim t'$ iff $[t]=[t']$}
\end{displaymath}
This follows from the above result of~\cite{rt} due to the Axiom
of Extensionality for $\Set$. 
\end{example}

\begin{observation}
\label{obs:3.9}
The category $\Class$ has all small limits and all class-indexed
colimits. That is, given a functor $D:\kat{D}\to\Class$ then
\begin{itemize}
\item[(a)] if $\kat{D}$ is small then $\lim D$ exists
\end{itemize}
and
\begin{itemize}
\item[(b)] if $\kat{D}$ has only a class of morphisms then
           $\colim D$ exists.
\end{itemize}
In fact, the strongly inaccessible cardinal $\Aleph$ which is
the cardinality of all classes, satisfies
\begin{displaymath}
(\Aleph)^n = \Aleph
\quad
\mbox{for all $n\in\Card$}
\end{displaymath}
Consequently, a cartesian product of a small collection of 
classes is a class --- thus, $\Class$ has small products.
And since small limits are always subobjects of small products,
it follows that $\Class$ has all smalll limits. Analogously, since 
\begin{displaymath}
\Aleph\o\Aleph = \Aleph
\end{displaymath}
it follows that $\Class$ has all class-indexed coproducts:
a disjoint union of a class of classes is a class. Since
class-indexed colimits are always quotients of class-indexed
coproducts, it follows that $\Class$ has class-indexed colimits.
\end{observation}

\begin{example}
Class-indexed limits do not exist, in general. For example, if $I$
is a proper class then $2^I$ (a cartesian product of $I$
copies of the two-element set $2=\{ 0,1\}$) is not a class, having
cardinality $2^\Aleph>\Aleph$. It follows that a product of $I$
copies of $2$ does not exist in the category $\Class$: it is trivial
that if $(\pi_i:L\to 2)_{i\in I}$ were such a product, then for 
every subclass $J\subseteq I$ we have the unique $u_J:1\to L$
with $\pi_i\o u_J$ given by $0$ for $i\in J$ and $1$ for 
$j\in I\setminus J$. Then the $u_J$'s are pairwise distinct,
thus, ${\mathrm{card}}\; L > \Aleph$, a contradiction.
\end{example}

Recall from~\cite{at} that an endofunctor $H$ of $\Class$
is called a {\em varietor} provided that every object of
$\Class$ generates a free $H$-algebra. Equivalently, if
the forgetful functor $\Alg{H}\to\Class$ has a left adjoint.

And an initial algebra of the functor 
\begin{displaymath}
H(\,\_\,)+A
\end{displaymath}
is precisely a free $H$-algebra on $H$. The former exists
by Corollary~\ref{cor:3.7}, thus we obtain

\begin{corollary}
Every endofunctor of $\Class$ is a varietor.
\end{corollary}

\begin{remark}
The category $\Alg{H}$ has all small limits and all class-indexed
colimits for every endofunctor $H$ of $\Class$: the limits are 
(obviously) created by the forgetful functor. The existence of colimits
follows from the fact that $({\mathrm{Epi}},{\mathrm{Mono}})$ is a 
factorization system in $\Class$ and every endofunctor $H$ preserves
epimorphisms (since they split). Since $H$ is a varietor, $\Alg{H}$
has all colimits which $\Class$ has, see~\cite{kelly}, Theorem~16.5
(where $\Alg{H}$ is shown to be reflective in the category 
$H/\Class$ having all colimits that $\Class$ has).
\end{remark}

\begin{odstavec}
{\bf Coalgebras}
of an endofunctor $H$ of $\Class$ are classes $A$ together 
with a function $\alpha:A\to HA$.
Given another algebra $\beta:B\to HB$, a {\em homomorphism} 
from $A$ to $B$ is a function $f:A\to B$ such that the following 
square
\begin{displaymath}
\miniSQUARE{A}
           {\alpha}
           {HA}
           {f}
           {Hf}
           {B}
           {\beta}
           {HB}
\end{displaymath}
commutes. The category of all $H$-coalgebras and homomorphisms 
is denoted by
\begin{displaymath}
\Coalg{H}
\end{displaymath}
\end{odstavec}

\begin{examples}
\mbox{}\hfill
\begin{enumerate}
\item Coalgebras over polynomial functors describe deterministic
      dynamic systems, see~\cite{r}. For example, if $\Sigma$ 
      consists of a binary symbol and a nullary one,
      then a coalgebra
      \begin{displaymath}
      A\to A\times A+1
      \end{displaymath}
      describes a system with the state-set $A$ and two deterministic
      inputs ($0$, $1$, say) with exceptions: to every state $a$
      the pair $(a_0,a_1)$ of states is assigned, representing
      the reaction of $a$ to $0$ and $1$, respectively --- unless
      $a$ is an exception, mapped to the unique element of $1$.
\item $\P^\infty$-coalgebras can be identified with large 
      small-branching graphs, i.e., classes $A$ endowed with 
      a binary relation (represented by the function
      $A\to\P^\infty A$ assigning to every node the small set
      of its descendants).
\end{enumerate}
\end{examples}

\begin{theorem}
\label{th:3.14}
Every endofunctor $H$ of $\Class$ has a terminal coalgebra.
\end{theorem}

Several proofs of this theorem are known. The first one is due
to Peter Aczel and Nax Mendler~\cite{am}. Their Final Coalgebra
Theorem states that every set-based endofunctor has a terminal
coalgebra --- but we know from Section~\ref{sec:2} that all 
endofunctors are set-based. Another proof follows, as Michael
Barr has noticed in~\cite{b}, from the theory of accessible categories
in the monograph~\cite{mp}. A third proof can be derived from
the result of James Worell~\cite{w} that every $\lambda$-accessible
endofunctor $H$ of $\Set$ has a terminal coalgebra obtained by
$2\lambda$ steps of the dual chain of~\ref{cor:3.7}(2). That is, define
a chain $V$ of (in general, large) sets as follows:
\begin{description}
\item[First step] $V_0=1$, $V_1=H(1)$ and
     $V_{0,1}:H(1)\to 1$ unique.
\item[Isolated step] $V_{i+1}=H(V_i)$ and 
     $V_{i+1,j+1}=H(V_{i,j})$.
\end{description}
and
\begin{description}
\item[Limit step] $V_j=\lim_{i<j} V_i$ with limit 
     cone $(V_{i,j})_{i<j}$
\end{description}
Then
\begin{displaymath}
\mbox{$V_{2\lambda}$ is a terminal coalgebra of $H$.}
\end{displaymath}
By applying this to $\lambda=\Aleph$ we ``almost'' obtain
a construction of terminal coalgebras of endofunctors of
$\Class$ (first, one has to extend the endofunctor to the
category of all large sets but this brings no difficulty). 
There is a catch here: although the resulting
limit $V_{2\Aleph}$ is indeed a class (which follows from
Worell's result), the intermediate step $V_\Aleph$ can
``slip'' outside the scope of classes:

\begin{example}
A terminal coalgebra $T$ of $\P^\infty$ has,
in the non-well-founded set theory of Peter Aczel, see~\cite{a}
or~\cite{bm}, a beautiful description: $T$ is the class
of all non-well-founded sets --- see~\cite{rt}. However, we work
here in the well-founded set theory ${\sf{ZFC}}$. An explicit
(but certainly not very beautiful) description of $T$ is 
presented in Section~\ref{sec:5} below.

Here we just observe that the chain $V$ above ``jumps'' out
of the realm of classes: if we put $V_0=1$, 
$V_{i+1}={\mathrm{exp}}\; V_i$ and $V_j=\lim_{i<j} V_i$
for all ordinals in $\Ord$, then we cannot form
\begin{displaymath}
V_\Aleph=\lim_{i\in\Ord} V_i
\end{displaymath}
within $\Class$.  The reason is that for all $i\leq\Aleph$ 
we can easily prove by transfinite induction that
\begin{displaymath}
{\mathrm{card}}\; V_i \geq 2^i
\end{displaymath}
Thus, $V_\Aleph$ is not a class.
\end{example}

\begin{remark}
In spite of the three proofs mentioned above, we present a new 
proof, based on ideas of Peter Gumm and Tobias Schr\"{o}der~\cite{gs}
since it is the shortest and clearest one, and it gives a sort of
concrete description: a terminal $H$-coalgebra is obtained from
a terminal $H_\Sigma$-coalgebra via a suitable congruence.
Recall that for every $H$-coalgebra $\alpha:A\to HA$ a {\em congruence}
is a quotient $e:A\to A/\sim$ of $A$ in $\Class$ for which a 
(necessarily unique)
structure map $\overline{\alpha}:A/\sim \to H(A/\sim)$ exists 
turning $e$ into a homomorphism:
\begin{displaymath}
\miniSQUARE{A}
           {\alpha}
           {HA}
           {e}
           {He}
           {A/\sim}
           {\overline{\alpha}}
           {H(A/\sim)}
\end{displaymath}
Recall further that a nice description of terminal $H_\Sigma$-coalgebras
is known, which works for large signatures as well as for small ones: let
\begin{displaymath}
T_\Sigma
\end{displaymath}
be the class of all (small) $\Sigma$-trees. (In comparison to $I_\Sigma$,
we just drop the well-foundedness.) This is, like $I_\Sigma$,
a $\Sigma$-algebra w.r.t. tree tupling --- and since in both cases
tree-tupling is actually an isomorphism we can invert it to
the structure map $\tau_\Sigma:T_\Sigma\to H_\Sigma T_\Sigma$
of a coalgebra. And that coalgebra is terminal.
\end{remark}

\begin{proposition}
\label{prop:3.17}
Every endofunctor $H$ of $\Class$, represented as a quotient
$\eps:H_\Sigma\to H$ (as in Proposition~\ref{prop:3.3}) has
a terminal coalgebra, viz, the quotient of the $H$-coalgebra
\begin{displaymath}
\xy
\xymatrix{
T_\Sigma
\ar[0,1]^-{\tau_\Sigma}
&
H_\Sigma T_\Sigma
\ar[0,1]^-{\eps_{T_\Sigma}}
&
HT_\Sigma
}
\endxy
\end{displaymath}
modulo the largest congruence.
\end{proposition}
\begin{proof}
\mbox{}\hfill
\begin{enumerate}
\item The largest congruence exists. In fact, the pushout of
      all congruences of $T_\Sigma$ is easily seen (due to
      the universal property of pushouts) to be a congruence.
\item Given the largest congruence $e:T_\Sigma\to T/\sim$,
      the corresponding coalgebra 
      $\overline{\tau_\Sigma}:T_\Sigma/\sim\to H(T_\Sigma/\sim)$
      is terminal. In fact, given a coalgebra $\beta:B\to HB$
      the uniqueness of a homomorphism from $B$ to $T_\Sigma/\sim$
      follows from the observation that given two homomorphisms
      $f_1,f_2:B\to T_\Sigma/\sim$, then a coequalizer
      $c:T_\Sigma/\sim\to T_\Sigma/\approx$ of $f_1$, $f_2$ 
      in $\Class$ yields a congruence $ce:T_\Sigma\to T_\Sigma/\approx$,
      thus, $\approx$ and $\sim$ coincide, which means $f_1=f_2$.
      The existence of a homomorphism is proved by choosing
      a splitting of the epimorphism $\eps_B$:
      \begin{displaymath}
      u:HB\to H_\Sigma B
      \quad
      \mbox{with $\eps_B u=\id$}
      \end{displaymath}
      The unique homomorphism, $f$, of the $H_\Sigma$-coalgebra
      $
      \xy
      \xymatrix@1{
      B
      \ar[0,1]^-{\beta}
      &
      HB
      \ar[0,1]^-{u}
      &
      H_\Sigma B
      }
      \endxy
      $ 
      yields a homomorphism, $ef$, of $H$-coalgebras:
      \begin{displaymath}
      \xy
      \xymatrix{
      &
      &
      &
      HB
      \ar[1,0]^{Hf}
      \\
      B
      \ar[0,1]^-{\beta}
      \ar[1,0]_f
      &
      HB
      \ar[0,1]_-{u}
      \ar @{=} [-1,2]
      &
      H_\Sigma B
      \ar[-1,1]_{\eps_B}
      \ar[1,0]^{H_\Sigma f}
      &
      HT_\Sigma
      \ar[2,0]^{He} 
      \\
      T_\Sigma
      \ar[0,2]_-{\tau_\eps}
      \ar[1,0]_{e}
      &
      &
      H_\Sigma T_\Sigma
      \ar[-1,1]_{\eps_{T_\Sigma}}
      &
      \\
      T_\Sigma/\sim
      \ar[0,3]_-{\overline{\tau_\Sigma}}
      &
      &
      &
      H(T_\Sigma/\sim)
      }
      \endxy
      \end{displaymath}
\end{enumerate}
\end{proof}

\begin{corollary}
Every endofunctor $H$ of $\Class$ is a covarietor, i.e., 
a cofree $H$-coalgebra on every class exists.
\end{corollary}

In fact, a cofree $H$-coalgebra on $A$ is just a terminal coalgebra
of $H(\,\_\,)\times A$.

\begin{remark}
The category $\Coalg{H}$ has all small limits and all class-indexed
colimits for every endofunctor $H$ of $\Class$: the colimits are 
(obviously) created by the forgetful functor. The existence of limits
follows, if $H$ preserves monomorphisms, from the dualization of
Theorem~16.5 of~\cite{kelly}. For general $H$ use the result of
V\v{e}ra Trnkov\'{a} cited in the proof of Theorem~\ref{th:2.3}
above.
\end{remark}

\section{All Functors Generate Completely Iterative Monads}
\label{sec:4}

In this section we assume that the reader is acquainted with
the concept of an iterative theory (or iterative monad)
of Calvin Elgot, and the coalgebraic treatment of completely 
iterative monads in~\cite{m} or~\cite{aamv}.
In~\cite{aamv} we worked with endofunctors $H$ such that 
a terminal coalgebra, $TX$, of the endofunctor $H(\,\_\,)+X$
exists for every $X$. Such functors were called {\em iteratable}.
In the category of classes this concept need not be used:

\begin{corollary}
Every endofunctor of $\Class$ is iteratable.
\end{corollary}

This follows from Proposition~\ref{prop:3.17} applied to $H(\,\_\,)+X$.

Recall from~\cite{m} or~\cite{aamv} that the coalgebra structure
of $TX$, 
$
\xy
\xymatrix@1{
TX
\ar[0,1]^-{\cong}
&
HTX+X
}
\endxy
$,
turns $TX$ into a coproduct of $HTX$ and $X$, where the coproduct
inclusions are denoted by
\begin{itemize}
\item[] $\tau_X:HTX\to TX$ ($TX$ is an $H$-algebra)
\end{itemize}
and
\begin{itemize}
\item[] $\eta_X:X\to TX$ ($X$ is contained in $TX$)
\end{itemize}
It turns out that this is part of a monad ${\mathbb{T}}=(T,\eta,\mu)$.
This monad is {\em completely iterative}, i.e., for every ``equation''
morphism $e:X\to T(X+Y)$ which is guarded, i.e., it factorizes
through the coproduct injection
\begin{displaymath}
HT(X+Y)+Y\subto HT(X+Y)+X+Y=T(X+Y)
\end{displaymath}
there exists a unique {\em solution}. That is, a unique morphism
$\sol{e}$ for which the following square
\begin{displaymath}
\xy
\xymatrix{
X
\ar[0,2]^-{\sol{e}}
\ar[1,0]_{e}
&
&
TY
\\
T(X+Y)
\ar[0,2]_-{T[\sol{e},\eta_Y]}
&
&
TTY
\ar[-1,0]_{\mu_Y}
}
\endxy
\end{displaymath}
commutes. And in~\cite{aamv} it has been proved that ${\mathbb{T}}$
can be characterized as a free completely iterative monad on $H$.

\begin{example}
Let $\Sigma$ be a (possibly large, infinitary) signature, i.e.,
a class of operation symbols together with a function ${\mathrm{ar}}(\,\_\,)$
assigning a small cardinal to every symbol $\sigma$. Put
$\Sigma_n=\{\sigma\mid {\mathrm{ar}}(\sigma)=n\}$. The polynomial
functor
\begin{displaymath}
H_\Sigma : X\mapsto \coprod_{\sigma\in\Sigma} X^{{\mathrm{ar}}(\sigma)}
\end{displaymath}
generates the following completely iterative monad ${\mathbb{T}}_\Sigma$:
\begin{itemize}
\item[] $T_\Sigma Y$ is the $\Sigma$-algebra of all $\Sigma$-trees
        on $Y$, i.e., small trees with leaves labelled in $\Sigma_0+Y$
        and nodes with $n>0$ children labelled in $\Sigma_n$,
\end{itemize}
and
\begin{itemize}
\item[] $\eta_Y$ is the singleton-tree embedding.
\end{itemize}
The fact that ${\mathbb{T}}_\Sigma$ is completely iterative just restates
the well-known property of tree algebras: all iterative systems
of equations that are guarded (i.e., do not contain equations $x\approx x'$
where $x$ and $x'$ are variables) have unique solutions.
\end{example}

\begin{corollary}
All free completely iterative monads on $\Class$ are quotient
monads of the tree-monads ${\mathbb{T}}_\Sigma$ (for all signatures
$\Sigma$).
\end{corollary}

In fact, every endofunctor $H$ of $\Class$ is a quotient of $H_\Sigma$
for a suitable signature $\Sigma$ (denote by $\Sigma_n$ the class $Hn$).
It follows that a free completely iterative monad on $H$ is 
a quotient of ${\mathbb{T}}_\Sigma$, see~\cite{a2}.

\section{Terminal Coalgebra of the Power-Set Functor}
\label{sec:5}

We apply the above results to non-labelled transition systems,
i.e., to coalgebras of the power-set functor $\P:\Set\to\Set$.
It has been noticed by several authors, e.g., \cite{am},
\cite{b}, \cite{jptww}, \cite{rt}, \cite{w} that $\P^{\infty}$ 
has a very natural {\em weakly terminal}
coalgebra $B$ (i.e., such that every $\P$-coalgebra $A$ has at
least one homomorphism from $A$ to $B$): the coalgebra of all
small extensional (see~\ref{ex:3.8}) trees. 
Throughout this section trees are always taken up to (graph)
isomorphism. Thus, shortly, a tree is extensional if and only if
distinct siblings define distinct subtrees. 

The weakly terminal coalgebra $B$ has as elements all small extensional 
trees, and the coalgebra structure
\begin{displaymath}
\beta:B\to \P^{\infty} B
\end{displaymath}
is the inverse of tree tupling, i.e., $\beta$ assigns to every
tree $t$ the set of all children of $t$.

We know from Theorem~\ref{th:3.14} that a terminal coalgebra for $\P^\infty$ 
exists. Since $B$ is weakly terminal, it follows that a terminal coalgebra
is a quotient of $B$ modulo the {\em bisimilarity equivalence}
$\bisim$ (i.e., the largest bisimulation on $B$). We are going
to describe this equivalence $\bisim$. We start by describing one
interesting class.

\begin{example}
\label{ex:Omega}
An extensional tree $t$ is bisimilar to the following tree
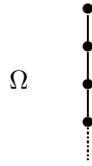
\begin{displaymath}
\Omega
\quad
\quad
\vcenter{
\xy
\POS (00,00) *{}           = "A"
   , (00,05) *{\bullet}    = "B"
   , (00,10) *{\bullet}    = "C"
   , (00,15) *{\bullet}    = "D"
   , (00,20) *{\bullet}    = "E"
\POS "A" \ar @{.} "B"
\POS "B" \ar @{-} "C"
\POS "C" \ar @{-} "D"
\POS "D" \ar @{-} "E"
\endxy
}
\end{displaymath}
if and only if all paths in $t$ are infinite. Thus, for example,
the following tree
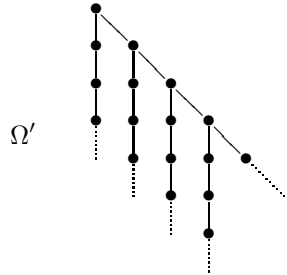
\begin{displaymath}
\Omega'
\quad
\quad
\vcenter{
\xy
\POS (00,00) *{\bullet}   = "A1"
   , (00,-05) *{\bullet}  = "B1"
   , (00,-10) *{\bullet}  = "C1"
   , (00,-15) *{\bullet}  = "D1"
   , (00,-20) *{}         = "E1"
   , (05,-05) *{\bullet}  = "A2"
   , (05,-10) *{\bullet}  = "B2"
   , (05,-15) *{\bullet}  = "C2"
   , (05,-20) *{\bullet}  = "D2"
   , (05,-25) *{}         = "E2"
   , (10,-10) *{\bullet}  = "A3"
   , (10,-15) *{\bullet}  = "B3"
   , (10,-20) *{\bullet}  = "C3"
   , (10,-25) *{\bullet}  = "D3"
   , (10,-30) *{}         = "E3"
   , (15,-15) *{\bullet}  = "A4"
   , (15,-20) *{\bullet}  = "B4"
   , (15,-25) *{\bullet}  = "C4"
   , (15,-30) *{\bullet}  = "D4"
   , (15,-35) *{}         = "E4"
   , (20,-20) *{\bullet}  = "A5"
   , (25,-25) *{}         = "A6"
\POS "A1" \ar @{-} "B1"
\POS "B1" \ar @{-} "C1"
\POS "C1" \ar @{-} "D1"
\POS "D1" \ar @{.} "E1"
\POS "A2" \ar @{-} "B2"
\POS "B2" \ar @{-} "C2"
\POS "C2" \ar @{-} "D2"
\POS "D2" \ar @{.} "E2"
\POS "A3" \ar @{-} "B3"
\POS "B3" \ar @{-} "C3"
\POS "C3" \ar @{-} "D3"
\POS "D3" \ar @{.} "E3"
\POS "A4" \ar @{-} "B4"
\POS "B4" \ar @{-} "C4"
\POS "C4" \ar @{-} "D4"
\POS "D4" \ar @{.} "E4"
\POS "A1" \ar @{-} "A2"
\POS "A2" \ar @{-} "A3"
\POS "A3" \ar @{-} "A4"
\POS "A4" \ar @{-} "A5"
\POS "A5" \ar @{.} "A6"
\endxy
}
\end{displaymath}
is bisimilar to $\Omega$. This illustrates that the
bisimilarity equivalence is non-trivial. We prove
$\Omega\bisim\Omega'$ below.
\end{example}

\begin{remark}
For the finite-power-set functor $\P_f$ a nice desription of
a terminal coalgebra has been presented by Michael Barr~\cite{b}:
let $B_f$ denote the coalgebra of all finitely branching extensional
trees. This is a small subcoalgebra of our (large) coalgebra $B$.
We call two trees $b$, $b'$ in $B_f$ {\em Barr-equivalent},
notation
\begin{displaymath}
b\rel{0} b'
\end{displaymath}
provided that for every natural number $n$ the tree $b|_n$ obtained
by cutting $b$ at level $n$ has the same extensional quotient
(sse~\ref{ex:3.8}) as the tree $b'|_n$. For example
\begin{displaymath}
\Omega\rel{0}\Omega'
\end{displaymath}
Barr proved that the quotient coalgebra
\begin{displaymath}
B_f/\!\rel{0}
\end{displaymath}
is a terminal $\P_f$-coalgebra --- that is, $\rel{0}$ is the bisimilarity
equivalence on $B_f$.
\end{remark}

We define, for every small ordinal number $i$, the
following equivalence relation $\rel{i}$ on $B$:
\begin{itemize}
\item[] $\rel{0}$ is the Barr-equivalence
\end{itemize}
and in case $i>0$
\begin{itemize}
\item[] $t\rel{i}s$ iff for all $j < i$ the following hold:
        \begin{itemize}
        \item[] \begin{itemize}
                \item[(1)] for each child $t'$ of $t$ there exists
                           a child $s'$ of $s$ such that $t' \rel{j} s'$
                \end{itemize}
        \end{itemize}
        and
        \begin{itemize}
        \item[] \begin{itemize}
                \item[(2)] vice versa.
                \end{itemize}
        \end{itemize}
\end{itemize}

\begin{remark}
We shall show below that the bisimilarity equivalence $\bisim$
is just the intersection of all $\rel{i}$. Notice that this
intersection is just the usual construction of a greatest
fixed point. Indeed, consider the collection $\Rel$ of all 
binary relations on $B$. This collection, ordered by set-inclusion,
is a class-complete lattice. Define $\Phi:\Rel\to\Rel$ as follows:
\begin{displaymath}
\begin{array}{rcl}
t\mathrel{\Phi(R)}s
&
\quad
\mbox{iff}
\quad
&
\mbox{for every child $t'$ of $t$ there exists a child $s'$}
\\
&
&
\mbox{of $s$ such that $t'\,R\,s'$, and vice versa.}
\end{array}
\end{displaymath}
Observe that $\Phi$ is a monotone function. Moreover, a binary
relation $R$ is a fixed point of $\Phi$ if and only if $R$ is
a bisimulation on $B$. Notice that the definition of $\rel{i}$
is just an iteration of $\Phi$ on the largest equivalence relation
$\approx_0$ (i.e., $B\times B$) shifted by $\omega$ steps: we have
\begin{displaymath}
{\rel{0}} = \Fi{\omega}(\approx_0)
\end{displaymath}
where for every relation $R$ the iterations $\Fi{i}(R)$, $i\in\Ord$,
are defined inductively as follows:
$\Fi{0}(R) = R$, the isolated step is $\Fi{i+1}(R) = \Phi(\Fi{i}(R))$,
and for limit ordinals $\Fi{i}(R) = \bigcap_{j<i} \Fi{j}(R)$.
Consequently, ${\rel{i}} = \Fi{\omega +i}(\approx_0)$.

That we are indeed constructing the largest fixed point for $\Phi$ 
follows from the following
\end{remark}

\begin{lemma}
\label{lem:fixed}
$\Phi$ preserves intersections of descending $\Ord$-chains.
\end{lemma}
\begin{proof}
Let $(R_i)_{i\in\Ord}$ be a descending chain in $\Rel$ and let
\begin{displaymath}
R = \bigcap_{i\in\Ord} R_i
\end{displaymath}
be its intersection. We show that $\Phi(R) = \bigcap_{i\in\Ord} \Phi(R_i)$.
In fact, the inclusion from left to right is obvious. To show the
inclusion from right to left, suppose that $t\mathrel{\Phi(R_i)} s$ 
holds for all $i\in\Ord$. Let $t'$ be any child of $t$.
Then, for any ordinal number $i\in\Ord$ there exists a child
$s_i'$ of $s$ with $t\mathrel{R_i}s_i'$.
Since $s$ has only a small set of children the set
$\{s_i' \mid i\in\Ord\}$ is small, too. Therefore there is a
cofinal subset $C$ of $\Ord$ such that $\{s_i' \mid i\in C\}$
has only one element, $s'$ say. It follows that $t'\mathrel{R_i}s'$
for all $i\in\Ord$. Hence, $t\mathrel{\Phi(R)} s$, as desired.
\end{proof}

\begin{theorem}
\label{th:5.5}
Two trees $t,s\in B$ are bisimilar iff $t\rel{i}s$ holds for
all small ordinals $i$.
\end{theorem}
\begin{proof}
It follows from Lemma \ref{lem:fixed} that the intersection
of all ${\rel{i}} = \Fi{i}(\rel{0})$, $i\in\Ord$ is a fixed point
of $\Phi$.

Next form the quotient coalgebra $B/\!\bisim$. Since $B$ is weakly terminal,
so is $B/\!\bisim$. In order to establish that $B/\!\bisim$ is a terminal
$\P^{\infty}$-coalgebra we must show that for any $\P^{\infty}$-coalgebra
$(X,\xi)$ and any two coalgebra homomorphisms $h, k:(X,\xi) \to (B,\beta)$
we have $h(x) \bisim k(x)$ for all $x \in X$. We show this by transfinite
induction. We write
\begin{displaymath}
\beta(k(x)) = \{s_j^x \mid j \in J_x\}
\quad
\mbox{and}
\quad
\beta(h(x)) = \{t_i^x \mid i \in I_x\}
\end{displaymath}
for the sets of children of $k(x)$ and $h(x)$, respectively. Since
$h$ and $k$ are coalgebra homomorphisms we have
\begin{displaymath}
\{s_j^x \mid j \in J_x\} = \{kx_\ell \mid \ell \in L_x\}
\quad
\mbox{and}
\quad
\{t_i^x \mid i \in I_x\} = \{hx_\ell \mid \ell \in L_x\},
\end{displaymath}
where $\xi(x) = \{x_\ell \mid \ell \in L_x\}$. 
\begin{description}
\item[First step, $i=0$]
     We will show that $k(x) \rel{0} h(x)$, i.e.,
     $E(k(x)|_n) = E(h(x)|_n)$ for all $n<\omega$ by
     induction on $n$. The statement is obvious for $n=0$.
     For the induction step observe that
     \begin{eqnarray*}
     \{E(s_j^x |_n) \mid j \in J_x\}
     &=&
     \{E(kx_\ell |_n) \mid \ell \in L_x\}
     \\
     &=&
     \{E(hx_\ell |_n) \mid \ell \in L_x\}
     \\
     &=&
     \{E(t_i^x|_n) \mid i \in I_x\}
     \end{eqnarray*}
     by the induction hypothesis. Hence, $E(k(x)|_{n+1})$ and
     $E(h(x)|_{n+1})$ have the same sets of children
     and therefore are equal.
\item[Induction step]
     Suppose now that $i>0$ is any ordinal number and that for
     all $x\in X$, $k(x)\rel{j} h(x)$ holds for all $j<i$.
     Consider any child $s'$ of $k(x)$, i.e., $s' = kx_\ell$ for
     some $x_\ell\in\xi(x)$. Then $t' = hx_\ell$ is a child of
     $h(x)$ such that $s' \rel{j} t'$ for all $j<i$.
\end{description}
Hence, we obtain $k(x)\rel{i} h(x)$ for all $i\in\Ord$,
which implies the desired result.
\end{proof}

\begin{remark}
Barr showed that $\rel{0}$ is the bisimilarity equivalence on
the set of finitely branching trees. However, it is not
a bisimulation on $B$. In order to see this notice
that is suffices to find trees that are in $\rel{0}$ but not
in $\rel{1}$. Consider the following trees
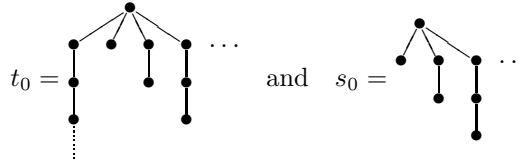
\begin{displaymath}
t_0=
\vcenter{
\xy
\POS (7.5,00) *{\bullet} = "A"
   , (00,-05) *{\bullet} = "B1"
   , (05,-05) *{\bullet} = "B2"
   , (10,-05) *{\bullet} = "B3"
   , (15,-05) *{\bullet} = "B4"
   , (00,-10) *{\bullet} = "C1"
   , (10,-10) *{\bullet} = "C3"
   , (15,-10) *{\bullet} = "C4"
   , (00,-15) *{\bullet} = "D1"
   , (15,-15) *{\bullet} = "D4"
   , (00,-20) *{}        = "E1"
   , (20,-05) *{\dots}
\POS "E1" \ar @{.} "D1"
\POS "D1" \ar @{-} "C1"
\POS "C1" \ar @{-} "B1"
\POS "B1" \ar @{-} "A"
\POS "B2" \ar @{-} "A"
\POS "C3" \ar @{-} "B3"
\POS "B3" \ar @{-} "A"
\POS "D4" \ar @{-} "C4"
\POS "C4" \ar @{-} "B4"
\POS "B4" \ar @{-} "A"
\endxy
}
\quad
\mbox{and}
\quad
s_0=
\vcenter{
\xy
\POS (7.5,00) *{\bullet} = "A"
   , (05,-05) *{\bullet} = "B2"
   , (10,-05) *{\bullet} = "B3"
   , (15,-05) *{\bullet} = "B4"
   , (10,-10) *{\bullet} = "C3"
   , (15,-10) *{\bullet} = "C4"
   , (15,-15) *{\bullet} = "D4"
   , (20,-05) *{\dots}
\POS "B2" \ar @{-} "A"
\POS "C3" \ar @{-} "B3"
\POS "B3" \ar @{-} "A"
\POS "D4" \ar @{-} "C4"
\POS "C4" \ar @{-} "B4"
\POS "B4" \ar @{-} "A"
\endxy
}
\end{displaymath}
We clearly have $t\rel{0} s$. But $t_0 \not\rel{1} s_0$,
since $t_0$ has a child which is an infinite path while
$s_0$ does not.
\end{remark}

\begin{definition}
We define trees $t_i$ and $s_i$ for all small ordinals $i$ for
which we show below that they are equivalent under $\rel{i}$
but not under $\rel{i+1}$.

\begin{enumerate}
\item We start with the trees $t_0$ and $s_0$ from the previous remark.
\item Given $t_i$ and $s_i$ we define
      \begin{displaymath}
      t_{i+1}=
      \vcenter
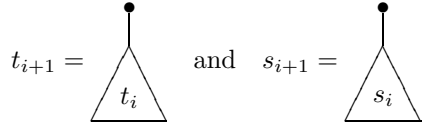
{
      \xy
      \POS (00,00) *{\bullet} = "A"
         , (00,-05) *{}       = "U"
         , (-5,-15) *{}       = "L"
         , (05,-15) *{}       = "R"
         , (00,-12) *{t_i}
      \POS "L" \ar @{-} "R"
      \POS "L" \ar @{-} "U"
      \POS "R" \ar @{-} "U"
      \POS "U" \ar @{-} "A"
      \endxy
      }
      \quad
      \mbox{and}
      \quad
      s_{i+1}=
      \vcenter{
      \xy
      \POS (00,00) *{\bullet} = "A"
         , (00,-05) *{}       = "U"
         , (-5,-15) *{}       = "L"
         , (05,-15) *{}       = "R"
         , (00,-12) *{s_i}
      \POS "L" \ar @{-} "R"
      \POS "L" \ar @{-} "U"
      \POS "R" \ar @{-} "U"
      \POS "U" \ar @{-} "A"
      \endxy
      }
      \end{displaymath}
\item For every limit ordinal $j$ we use the following auxilliary trees
      (where $i<j$ is arbitrary)
      \begin{displaymath}
      u_j=
      \vcenter
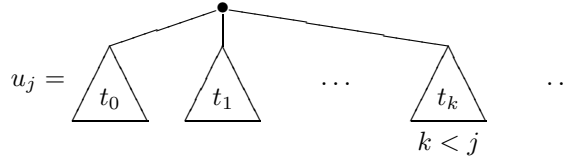
{
      \xy
      \POS (000,000) *{\bullet} = "A"
         , (-15,-05) *{}       = "U0"
         , (-20,-15) *{}       = "L0"
         , (-10,-15) *{}       = "R0"
         , (-15,-12) *{t_0} 
         , (000,-05) *{}       = "U1"
         , (-05,-15) *{}       = "L1"
         , (005,-15) *{}       = "R1"
         , (000,-12) *{t_1}
         , (015,-10) *{\dots}
         , (030,-05) *{}       = "Uk"
         , (025,-15) *{}       = "Lk"
         , (035,-15) *{}       = "Rk"
         , (030,-12) *{t_k}
         , (030,-18) *{k<j}
         , (045,-10) *{\dots}
      \POS "L0" \ar @{-} "R0"
      \POS "L0" \ar @{-} "U0"
      \POS "R0" \ar @{-} "U0"
      \POS "U0" \ar @{-} "A"
      \POS "L1" \ar @{-} "R1"
      \POS "L1" \ar @{-} "U1"
      \POS "R1" \ar @{-} "U1"
      \POS "U1" \ar @{-} "A"
      \POS "Lk" \ar @{-} "Rk"
      \POS "Lk" \ar @{-} "Uk"
      \POS "Rk" \ar @{-} "Uk"
      \POS "Uk" \ar @{-} "A"
      \endxy
      }
      \end{displaymath}
      and
      \begin{displaymath}
      v^i_j=
      \vcenter
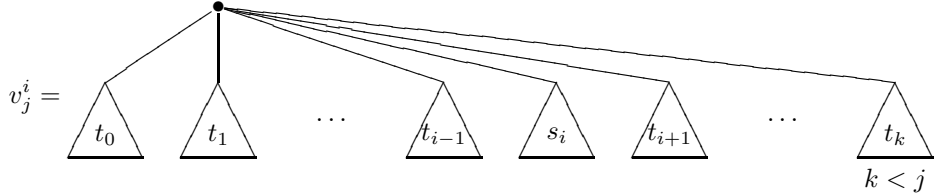
{
      \xy
      \POS (000,005) *{\bullet} = "A"
         , (-15,-05) *{}       = "U0"
         , (-20,-15) *{}       = "L0"
         , (-10,-15) *{}       = "R0"
         , (-15,-12) *{t_0} 
         , (000,-05) *{}       = "U1"
         , (-05,-15) *{}       = "L1"
         , (005,-15) *{}       = "R1"
         , (000,-12) *{t_1}
         , (015,-10) *{\dots}
         , (030,-05) *{}       = "Ui-1"
         , (025,-15) *{}       = "Li-1"
         , (035,-15) *{}       = "Ri-1"
         , (030,-12) *{t_{i-1}}
         , (045,-05) *{}       = "Ui"
         , (040,-15) *{}       = "Li"
         , (050,-15) *{}       = "Ri"
         , (045,-12) *{s_i}
         , (060,-05) *{}       = "Ui+1"
         , (055,-15) *{}       = "Li+1"
         , (065,-15) *{}       = "Ri+1"
         , (060,-12) *{t_{i+1}}
         , (075,-10) *{\dots}
         , (090,-05) *{}       = "Uk"
         , (085,-15) *{}       = "Lk"
         , (095,-15) *{}       = "Rk"
         , (090,-12) *{t_k}
         , (090,-18) *{k<j}         
      \POS "L0" \ar @{-} "R0"
      \POS "L0" \ar @{-} "U0"
      \POS "R0" \ar @{-} "U0"
      \POS "U0" \ar @{-} "A"
      \POS "L1" \ar @{-} "R1"
      \POS "L1" \ar @{-} "U1"
      \POS "R1" \ar @{-} "U1"
      \POS "U1" \ar @{-} "A"
      \POS "Li-1" \ar @{-} "Ri-1"
      \POS "Li-1" \ar @{-} "Ui-1"
      \POS "Ri-1" \ar @{-} "Ui-1"
      \POS "Ui-1" \ar @{-} "A"
      \POS "Li" \ar @{-} "Ri"
      \POS "Li" \ar @{-} "Ui"
      \POS "Ri" \ar @{-} "Ui"
      \POS "Ui" \ar @{-} "A"
      \POS "Li+1" \ar @{-} "Ri+1"
      \POS "Li+1" \ar @{-} "Ui+1"
      \POS "Ri+1" \ar @{-} "Ui+1"
      \POS "Ui+1" \ar @{-} "A"
      \POS "Lk" \ar @{-} "Rk"
      \POS "Lk" \ar @{-} "Uk"
      \POS "Rk" \ar @{-} "Uk"
      \POS "Uk" \ar @{-} "A"
      \endxy
      }
      \end{displaymath}
      and we define
      \begin{displaymath}
      t_j=
      \vcenter
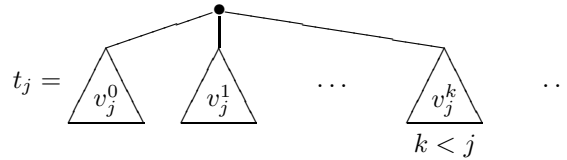
{
      \xy
      \POS (000,000) *{\bullet} = "A"
         , (-15,-05) *{}       = "U0"
         , (-20,-15) *{}       = "L0"
         , (-10,-15) *{}       = "R0"
         , (-15,-12) *{v^0_j} 
         , (000,-05) *{}       = "U1"
         , (-05,-15) *{}       = "L1"
         , (005,-15) *{}       = "R1"
         , (000,-12) *{v^1_j}
         , (015,-10) *{\dots}
         , (030,-05) *{}       = "Uk"
         , (025,-15) *{}       = "Lk"
         , (035,-15) *{}       = "Rk"
         , (030,-12) *{v^k_j}
         , (030,-18) *{k<j}
         , (045,-10) *{\dots}
      \POS "L0" \ar @{-} "R0"
      \POS "L0" \ar @{-} "U0"
      \POS "R0" \ar @{-} "U0"
      \POS "U0" \ar @{-} "A"
      \POS "L1" \ar @{-} "R1"
      \POS "L1" \ar @{-} "U1"
      \POS "R1" \ar @{-} "U1"
      \POS "U1" \ar @{-} "A"
      \POS "Lk" \ar @{-} "Rk"
      \POS "Lk" \ar @{-} "Uk"
      \POS "Rk" \ar @{-} "Uk"
      \POS "Uk" \ar @{-} "A"
      \endxy
      }
      \end{displaymath}
      and
      \begin{displaymath}
      s_j=
      \vcenter
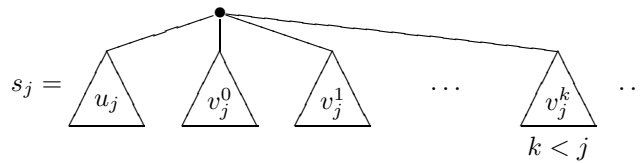
{
      \xy
      \POS (000,000) *{\bullet} = "A"
         , (-15,-05) *{}       = "U0"
         , (-20,-15) *{}       = "L0"
         , (-10,-15) *{}       = "R0"
         , (-15,-12) *{u_j} 
         , (000,-05) *{}       = "U1"
         , (-05,-15) *{}       = "L1"
         , (005,-15) *{}       = "R1"
         , (000,-12) *{v^0_j}
         , (015,-05) *{}       = "U2"
         , (010,-15) *{}       = "L2"
         , (020,-15) *{}       = "R2"
         , (015,-12) *{v^1_j}
         , (030,-10) *{\dots}
         , (045,-05) *{}       = "Uk"
         , (040,-15) *{}       = "Lk"
         , (050,-15) *{}       = "Rk"
         , (045,-12) *{v^k_j}
         , (045,-18) *{k<j}
         , (055,-10) *{\dots}
      \POS "L0" \ar @{-} "R0"
      \POS "L0" \ar @{-} "U0"
      \POS "R0" \ar @{-} "U0"
      \POS "U0" \ar @{-} "A"
      \POS "L1" \ar @{-} "R1"
      \POS "L1" \ar @{-} "U1"
      \POS "R1" \ar @{-} "U1"
      \POS "U1" \ar @{-} "A"
      \POS "L2" \ar @{-} "R2"
      \POS "L2" \ar @{-} "U2"
      \POS "R2" \ar @{-} "U2"
      \POS "U2" \ar @{-} "A"
      \POS "Lk" \ar @{-} "Rk"
      \POS "Lk" \ar @{-} "Uk"
      \POS "Rk" \ar @{-} "Uk"
      \POS "Uk" \ar @{-} "A"
      \endxy
      }
      \end{displaymath}
\end{enumerate}
\end{definition}

\begin{theorem}
None of the equivalences $\rel{i}$ is a congruence.
\end{theorem}
\begin{proof}
We prove $t_i\rel{i} s_i$ but $t_i\not\rel{i+1} s_i$ for all ordinals $i$.
This proves the theorem.
\begin{enumerate}
\item Proof of $t_i\rel{i} s_i$. We proceed by transfinite induction 
      on $i$.
      \begin{description}
      \item[Initial case] $t_0\rel{0} s_0$ --- clear.
      \item[Isolated case] $t_i\rel{i} s_i$ clearly implies 
           $t_{i+1}\rel{i+1} s_{i+1}$.
      \item[Limit case] Let $j$ be a limit ordinal with $t_i\rel{i} s_i$
           for all $i<j$. Then, obviously, $u_j\rel{i} v^i_j$ for all
           $i<j$, thus, $u_j\rel{j} v^i_j$, which implies $t_j\rel{j} s_j$.
      \end{description}
\item 
\label{item:2}
      We need some auxilliary facts about cuttings $w|_n$ of trees
      $w$ at level $n$:
      \begin{itemize}
      \item[(a)] For $n=1$ all the trees $t_i$, $s_i$, $u_j$, $v^i_j$
                 cut to
                 \begin{displaymath}
                 \xy
                 \POS (00,00) *{\bullet} = "A"
                    , (00,-5) *{\bullet} = "B"
                 \POS "A" \ar @{-} "B"
                 \endxy
                 \end{displaymath}
                 because they have all more than one vertex --- this
                 is obvious.
      \item[(b)] We have
                 \begin{displaymath}
                 t_0|_2 = s_0|_2 =
                 \vcenter
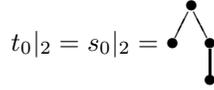
{
                 \xy
                 \POS (000,000) *{\bullet} = "A"
                    , (000,-05) *{\bullet} = "B"
                    , (-05,000) *{\bullet} = "C"
                    , (-2.5,05) *{\bullet} = "D"
                 \POS "A" \ar @{-} "B"
                 \POS "A" \ar @{-} "D"
                 \POS "C" \ar @{-} "D"
                 \endxy
                 }
                 \end{displaymath}
                 and
                 \begin{displaymath}
                 t_i|_2 = s_i|_2 =
                 \vcenter
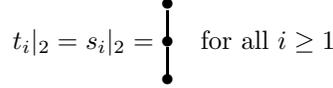
{
                 \xy
                 \POS (000,000) *{\bullet} = "A"
                    , (000,-05) *{\bullet} = "B"
                    , (000,-10) *{\bullet} = "C"
                 \POS "A" \ar @{-} "B"
                 \POS "B" \ar @{-} "C"
                 \endxy
                 }
                 \quad
                 \mbox{for all $i\geq 1$}
                 \end{displaymath} 
                 The first statement is obvious, and so is the second one
                 for isolated ordinals $i$. For limit ordinals it follows
                 from~(a).
      \item[(c)] We have
                 \begin{displaymath}
                 u_j|_2 = v^i_j|_2 =
                 \vcenter
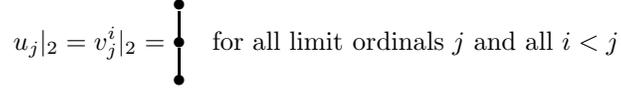
{
                 \xy
                 \POS (000,000) *{\bullet} = "A"
                    , (000,-05) *{\bullet} = "B"
                    , (000,-10) *{\bullet} = "C"
                 \POS "A" \ar @{-} "B"
                 \POS "B" \ar @{-} "C"
                 \endxy
                 }
                 \quad
                 \mbox{for all limit ordinals $j$ and all $i<j$}
                 \end{displaymath}                           
                 This follows from~(b).
      \item[(d)] We have
                 \begin{displaymath}
                 t_0|_3 = s_0|_3 =
                 \vcenter
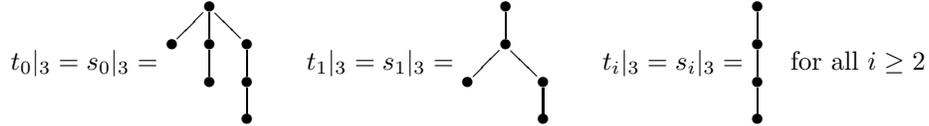
{
                 \xy
                 \POS (000,000) *{\bullet} = "A"
                    , (-05,-05) *{\bullet} = "A1"
                    , (000,-05) *{\bullet} = "B1"
                    , (000,-10) *{\bullet} = "B2"
                    , (005,-05) *{\bullet} = "C1"
                    , (005,-10) *{\bullet} = "C2"
                    , (005,-15) *{\bullet} = "C3"
                 \POS "A" \ar @{-} "A1"
                 \POS "A" \ar @{-} "B1"
                 \POS "B1" \ar @{-} "B2"
                 \POS "A" \ar @{-} "C1"
                 \POS "C1" \ar @{-} "C2"
                 \POS "C2" \ar @{-} "C3"
                 \endxy
                 }
                 \quad
                 \quad
                 t_1|_3 = s_1|_3 =
                 \vcenter{
                 \xy
                 \POS (000,000) *{\bullet} = "A"
                    , (000,-05) *{\bullet} = "A1"
                    , (-05,-10) *{\bullet} = "B1"
                    , (005,-10) *{\bullet} = "C1"
                    , (005,-15) *{\bullet} = "D"
                 \POS "A" \ar @{-} "A1"
                 \POS "A1" \ar @{-} "B1"
                 \POS "A1" \ar @{-} "C1"
                 \POS "C1" \ar @{-} "D"
                 \endxy
                 }
                 \quad
                 \quad
                 t_i|_3 = s_i|_3 =
                 \vcenter{
                 \xy
                 \POS (000,000) *{\bullet} = "A"
                    , (000,-05) *{\bullet} = "B"
                    , (000,-10) *{\bullet} = "C"
                    , (000,-15) *{\bullet} = "D"
                 \POS "A" \ar @{-} "B"
                 \POS "B" \ar @{-} "C"
                 \POS "C" \ar @{-} "D"
                 \endxy
                 }
                 \quad
                 \mbox{for all $i\geq 2$}
                 \end{displaymath} 
                 The last statement follows from~(c).
      \item[(e)] We have 
                 \begin{displaymath}
                 u_j|_3 = v^i_j|_3 =
                 \vcenter
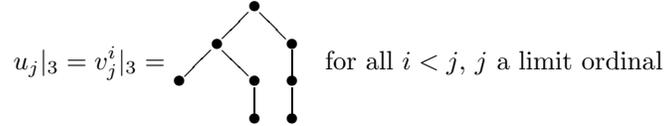
{
                 \xy
                 \POS (000,000) *{\bullet} = "A"
                    , (-05,-05) *{\bullet} = "B1"
                    , (005,-05) *{\bullet} = "B2"
                    , (-10,-10) *{\bullet} = "C1"
                    , (000,-10) *{\bullet} = "C2"
                    , (005,-10) *{\bullet} = "C3"
                    , (000,-15) *{\bullet} = "D1"
                    , (005,-15) *{\bullet} = "D2"
                 \POS "A" \ar @{-} "B1"
                 \POS "A" \ar @{-} "B2"
                 \POS "B1" \ar @{-} "C1"
                 \POS "B1" \ar @{-} "C2"
                 \POS "B2" \ar @{-} "C3"
                 \POS "C2" \ar @{-} "D1"
                 \POS "C3" \ar @{-} "D2"
                 \endxy
                 }
                 \quad
                 \mbox{for all $i<j$, $j$ a limit ordinal}
                 \end{displaymath}            
                 This follows from~(b).
      \item[(f)] We have
                 \begin{displaymath}
                 t_0|_4 = s_0|_4 =
                 \vcenter
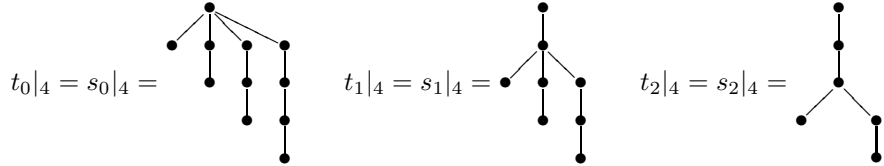
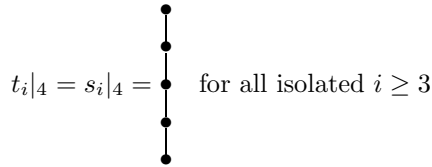
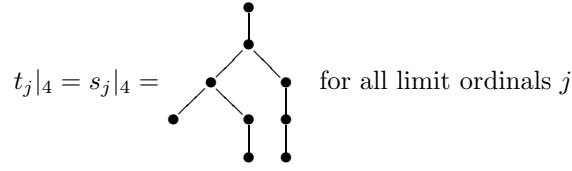
{
                 \xy
                 \POS (000,000) *{\bullet} = "A"
                    , (-05,-05) *{\bullet} = "B1"
                    , (000,-05) *{\bullet} = "B2"
                    , (005,-05) *{\bullet} = "B3"
                    , (010,-05) *{\bullet} = "B4"
                    , (000,-10) *{\bullet} = "C1"
                    , (005,-10) *{\bullet} = "C2"
                    , (010,-10) *{\bullet} = "C3"
                    , (005,-15) *{\bullet} = "D1"
                    , (010,-15) *{\bullet} = "D2"
                    , (010,-20) *{\bullet} = "E1"
                 \POS "A" \ar @{-} "B1"
                 \POS "A" \ar @{-} "B2"
                 \POS "A" \ar @{-} "B3"
                 \POS "A" \ar @{-} "B4"
                 \POS "B2" \ar @{-} "C1"
                 \POS "B3" \ar @{-} "C2"
                 \POS "B4" \ar @{-} "C3"
                 \POS "C2" \ar @{-} "D1"
                 \POS "C3" \ar @{-} "D2"
                 \POS "D2" \ar @{-} "E1"
                 \endxy
                 }
                 \quad
                 \quad
                 t_1|_4 = s_1|_4 =
                 \vcenter{
                 \xy
                 \POS (000,000) *{\bullet} = "A"
                    , (000,-05) *{\bullet} = "B"
                    , (-05,-10) *{\bullet} = "C1"
                    , (000,-10) *{\bullet} = "C2"
                    , (005,-10) *{\bullet} = "C3"
                    , (000,-15) *{\bullet} = "D1"
                    , (005,-15) *{\bullet} = "D2"
                    , (005,-20) *{\bullet} = "E1"
                 \POS "A" \ar @{-} "B"
                 \POS "B" \ar @{-} "C1"
                 \POS "B" \ar @{-} "C2"
                 \POS "B" \ar @{-} "C3"
                 \POS "C2" \ar @{-} "D1"
                 \POS "C3" \ar @{-} "D2"
                 \POS "D2" \ar @{-} "E1"
                 \endxy
                 }
                 \quad
                 \quad
                 t_2|_4 = s_2|_4 =
                 \vcenter{
                 \xy
                 \POS (000,000) *{\bullet} = "A"
                    , (000,-05) *{\bullet} = "B"
                    , (000,-10) *{\bullet} = "C"
                    , (-05,-15) *{\bullet} = "D1"
                    , (005,-15) *{\bullet} = "D2"
                    , (005,-20) *{\bullet} = "E1"
                 \POS "A" \ar @{-} "B"
                 \POS "B" \ar @{-} "C"
                 \POS "C" \ar @{-} "D1"
                 \POS "C" \ar @{-} "D2"
                 \POS "D2" \ar @{-} "E1"
                 \endxy
                 }
                 \end{displaymath}  
                 and
                 \begin{displaymath}
                 t_i|_4 = s_i|_4 =
                 \vcenter{
                 \xy
                 \POS (000,000) *{\bullet} = "A"
                    , (000,-05) *{\bullet} = "B"
                    , (000,-10) *{\bullet} = "C"
                    , (000,-15) *{\bullet} = "D"
                    , (000,-20) *{\bullet} = "E"
                 \POS "A" \ar @{-} "B"
                 \POS "B" \ar @{-} "C"
                 \POS "C" \ar @{-} "D"
                 \POS "D" \ar @{-} "E"
                 \endxy
                 }
                 \quad
                 \mbox{for all isolated $i\geq 3$}
                 \end{displaymath}
                 as well as
                 \begin{displaymath}
                 t_j|_4=s_j|_4=
                 \vcenter{
                 \xy
                 \POS (000,000) *{\bullet} = "A"
                    , (000,-05) *{\bullet} = "B"
                    , (-05,-10) *{\bullet} = "C1"
                    , (005,-10) *{\bullet} = "C2"                      
                    , (-10,-15) *{\bullet} = "D1"
                    , (000,-15) *{\bullet} = "D2"
                    , (005,-15) *{\bullet} = "D3"
                    , (000,-20) *{\bullet} = "E1"
                    , (005,-20) *{\bullet} = "E2"
                 \POS "A" \ar @{-} "B"
                 \POS "B" \ar @{-} "C1"
                 \POS "B" \ar @{-} "C2"
                 \POS "C1" \ar @{-} "D1"
                 \POS "C1" \ar @{-} "D2"
                 \POS "C2" \ar @{-} "D3"
                 \POS "D2" \ar @{-} "E1"
                 \POS "D3" \ar @{-} "E2"
                 \endxy
                 }                 
                 \quad
                 \mbox{for all limit ordinals $j$}
                 \end{displaymath}         
                 The last statement follows from~(e), the last but
                 one from~(d).
      \end{itemize}
\item 
\label{item:3}
      We prove
      \begin{displaymath}
      t_i\not\rel{i+2} t_k
      \quad
      \mbox{and}
      \quad
      s_i\not\rel{i+2} t_k
      \quad
      \mbox{for all ordinals $i<k$}
      \end{displaymath}
      We proceed by transfinite induction on $k$:
      \begin{itemize}
      \item[1.]  Initial case: there is nothing to prove if $k=0$.
      \item[2.]  Isolated case: $t_i\not\rel{i+2} t_{k+1}$ is clear
                 if $i=0$ (in fact, $t_0\not\rel{0} t_{k+1}$ because
                 $t_0|_2\not= t_{k_1}|_2$, see~(\ref{item:2}b)
                 and if $i$ is a limit ordinal $(t_i\not\rel{0} t_{k+1}$
                 because $t_i|_4\not= t_{k+1}|_4$, see~(\ref{item:2}f).
                 If $i$ is an isolated ordinal, then 
                 $t_{i-1}\not\rel{i+1} t_k$ implies $t_i\not\rel{i+2} t_{k+1}$.
                 Analogously with $s_i\not\rel{i+2} t_k$.
      \item[3.]  Limit case: let $k$ be a limit ordinal. We proceed 
                 by transfinite induction on $i$.
                 \begin{itemize}
                 \item[3.1] Initial case: $t_0\not\rel{2} t_k$ because
                            $t_0|_2\not= t_k|_2$, see~(\ref{item:2}b).
                            Analogously $s_0\not\rel{2} t_k$.
                 \item[3.2] Isolated case: $t_{i+1}\not\rel{i+3} t_k$
                            because
                            $t_{i+1}|_4\not= t_k|_4$, see~(\ref{item:2}f).
                            Analogously $s_{i+1}\not\rel{i+3} t_k$.
                 \item[3.3] Limit case: let $j<k$ be a limit ordinal.
                            Assuming $t_j\not\rel{j+2} t_k$, we derive
                            a contradiction. The child $u_k$ of $t_k$
                            must be $\rel{j+1}$-equivalent to a child 
                            of $t_j$, i.e.,
                            \begin{itemize}
                            \item[] either $u_j\rel{j+1} u_k$, or
                                    $v^i_j\rel{j+1} u_k$ for some $i<j$.
                            \end{itemize}  
                            The first possibility implies that the child 
                            $t_j$ of $u_k$ is $\rel{j}$-equivalent to a 
                            child $t_l$ of $u_j$, $l<j$. Thus, we have
                            \begin{itemize}
                            \item[] $t_l\rel{j} t_j$ for $l<j<k$.
                            \end{itemize}
                            This contradicts to the fact that, by induction,
                            $t_l\not\rel{l+2} t_j$ (and $l+2<j$).
                            Analogously with the second possibility,
                            $v^i_j\rel{j+1} u_k$, where the only case
                            that we have to consider extra is the child
                            $s_i$ of $v^i_j$ --- however,
                            \begin{displaymath}
                            s_i\rel{j} t_j
                            \end{displaymath}
                            is also a contradiction since, by induction,
                            $s_i\not\rel{i+2} t_j$ (and $l+2<j$).

                            Finally, assuming $s_j\rel{j+2} t_k$, we
                            derive a contradiction analogously, the only
                            new case to consider here is that the child
                            $t_j$ of $t_k$ is $\rel{j}$-equivalent
                            to the child $u_j$ of $s_j$:
                            \begin{displaymath}
                            u_j\rel{j} t_j
                            \end{displaymath}
                            This, however, is a contradiction again: we 
                            have $u_j\not\rel{0} t_j$ because
                            $u_j|_3\not= t_j|_3$, see~(\ref{item:2}d,e).
                 \end{itemize}
      \end{itemize}
\item Proof of $t_i\not\rel{i+1} s_i$. We proceed by transfinite 
      induction on $i$. 
      \begin{description}
      \item[Initial case] $t_0\not\rel{1} s_0$ by our choice of trees
           $t_0$ and $s_0$.
      \item[Isolated case] From $t_i\not\rel{i+1} s_i$ it follows
           immediately that $t_{i+1}\not\rel{i+2} s_{i+1}$.
      \item[Limit case] Let $j$ be a limit ordinal with $t_j\rel{j+1} s_j$.
           We derive a contradiction. The child $u_j$ of $s_j$ is 
           $\rel{0}$-equivalent to a child of $t_j$. That is, 
           \begin{itemize}
           \item[] $u_j\rel{j} v^k_j$ for some $k<j$.
           \end{itemize} 
           This implies that the child $t_k$ of $u_j$ is 
           $\rel{k+2}$-equivalent to some child of $v^k_j$, i.e., 
           \begin{itemize}
           \item[] either $t_k\rel{k+2} s_k$ or $t_k\rel{k+2} t_l$
                   for some $l\not= k$, $l<j$.
           \end{itemize}
           The first case does not happen: by induction hypothesis,
           $t_k\not\rel{k+1} s_k$. The second case contradicts 
           to~(\ref{item:3}): if $k<l$, and for $l<k$ we know 
           from~(\ref{item:3}) that $t_l\not\rel{l+2} t_k$, thus,
           again $t_l\not\rel{k+2} t_k$.
      \end{description}
\end{enumerate}
\end{proof}

\begin{remark}
We have described a terminal coalgebra of $\P^\infty$ as the coalgebra
of all extensional trees modulo the congruence $\bigcap_{i\in\Ord}\rel{i}$.
Since none of the equivalences is a congruence, we see no hope in obtaining
a nicer description of a terminal $\P^\infty$-coalgebra in well-founded
set theory.
\end{remark}


\begin{thebibliography}{MMMMM}
\bibitem[A]{a}
        P.~Aczel,
        {\em Non-Well-Founded Sets}, CSLI Lecture Notes 14,
        Stanford University, 1988
\bibitem[AM]{am}
        P.~Aczel and N.~Mendler, 
        A Final Coalgebra Theorem, 
        in: {\em Category Theory and Computer Science}, 
        D.~H. Pitt, D.~E. Rydebeard, P.~Dyjber, A.~M. Pitts, A.~Poign\'{e} 
        (eds.), LNCS~389, Springer-Verlag, 1989, 357--365
\bibitem[AAMV]{aamv}
        P.~Aczel, J.~Ad\'{a}mek, S.~Milius and J.~Velebil,
        Infinite Trees and Completely Iterative Theories:
        A Coalgebraic View,
        accepted for publication in 
        {\em Theor. Comp. Science} 
\bibitem[A${}_1$]{a1}
        J.~Ad\'{a}mek,
        Free Algebras and Automata Realization in the Language of 
        Categories,
        {\em Comm. Math. Univ. Carolinae} 15 (1974), 589--602
\bibitem[A${}_2$]{a2}
        J.~Ad\'{a}mek,
        A Description of Free Iterative Theories,
        {\em manuscript}        
\bibitem[AMV]{amv}
        J.~Ad\'{a}mek, S.~Milius and J.~Velebil,
        Final Coalgebras And a Solution Theorem For Arbitrary Endofunctors,
        {\em El. Notes Theor. Comp. Science} 65.1 (2002)
\bibitem[AP${}_1$]{ap1}
        J.~Ad\'{a}mek and H.-E.~Porst,
        On Tree Coalgebras and Coalgebra Presentations,
        {\em preprint}
\bibitem[AP${}_2$]{ap2}
        J.~Ad\'{a}mek and H.-E.~Porst,
        On Varieties and Covarieties in a Category,
        accepted for publication in {\em Math. Str. Comp. Science}
\bibitem[AT]{at}
        J.~Ad\'{a}mek and V.~Trnkov\'{a},
        {\em Automata and Algebras in a Category}, 
        Kluwer Publishing Company, 1990 
\bibitem[B]{b}
        M.~Barr, Terminal Coalgebras in Well-Founded Set Theory,
        {\em Theor. Comp. Science} 124 (1994), 182--192
\bibitem[BM]{bm}
        J.~Barwise and L.~Moss,
        {\em Vicious Circles},
        CSLI Lecture Notes No.~60,
        1996
\bibitem[GS]{gs}
        H.~P.~Gumm and T.~Schr\"{o}der,
        Coalgebras of bounded type, 
        {\em Math. Str. Comp. Science} 12 (2002), 565--578. 
\bibitem[JPTWW]{jptww}
        P.~Johnstone, J.~Power, T.~Tsujishita, H.~Watanabe
        and J.~Worell,
        On the Structure of Categories of Coalgebras,
        {\em Theor. Comp. Science} 260 (2001), 87--117
\bibitem[Ke]{kelly}
        G.~M.~Kelly,
        A unified treatment of transfinite constructions for free algebras, 
        free monoids, colimits, associated sheaves, and so on, 
        {\em Bulletin of the Australian Mathematical Society} 22 (1980), 1--83 
\bibitem[Ko]{k}
        V.~Koubek,
        Set Functors II --- Contaravariant Case,
        {\em Comm. Math. Univ. Carolinae} 14 (1973), 47--59
\bibitem[MP]{mp}
        M.~Makkai and R.~Par\'{e},
        {\em Accessible Categories: The Foundations of Categorical
        Model Theory},
        Contemporary Math., Providence, 1989
\bibitem[M]{m}
        L.~Moss,
        Parametric Corecursion,
        {\em Theor. Comput. Sci.} 260(1--2) (2001), 139--163
\bibitem[R]{r}
        J.~J.~M.~M.~Rutten,
        Universal coalgebra: a theory of systems,
        {\em Theoret. Comput. Science} 249(1) (2000), 3--80
\bibitem[RT]{rt}
        J.~J.~M.~M.~Rutten and D.~Turi,
        On the foundations of final coalgebra semantics:
        non-well-founded sets, partial orders, metric spaces,
        {\em Math. Struct. Comp. Science} Vol. 8 (1998), 481--540
\bibitem[T]{t}
        A.~Tarski,
        Sur la d\'{e}composition des ensembles en sous-ensembles
        pr\`{e}sque disjoint,
        {\em Fund. Math.} 14 (1929), 205--215
\bibitem[W]{w}
        J.~Worrell,
        {\em On Coalgebras and Final Semantics}, PhD thesis,
        Oxford Unversity Computing Laboratory, 2000
\end{thebibliography}
\end{document}